\documentclass[final]{eps}

\usepackage{graphicx}
\usepackage{times}
\usepackage{natbib}
\volumenumber{58}
\publishyear{2006}
\frompage{\pageref{firstpage}}
\topage{\pageref{finalpage}}

\shorttitle{A. Pollo \lowercase{\textit{et al.}}: Clustering of AKARI galaxies.}

\title{Clustering of Far-Infrared Galaxies in the AKARI All-Sky Survey}
\author{A. Pollo$^{1,2,3}$, T. T. Takeuchi$^{4}$, T. L. Suzuki$^4$, and S. Oyabu$^4$}
\affiliation{$^1$Center for Theoretical Physics of the Polish Academy of Sciences, Al. Lotnikow 32/46, 02-668 Warsaw, Poland\\
             $^2$Astronomical Observatory of the Jagiellonian University, Orla 171, 30-001 Cracow, Poland\\
             $^3$The Andrzej Soltan Institute for Nuclear Studies, Ho\.{z}a 69, 00-681 Warsaw, Poland\\
             $^4$Division of Particle and Astrophysical Science, Nagoya University, Furo-cho, Chikusa-ku, Nagoya 464-8602, Japan\\
             }
\received{xxxx xx, 2003}\revised{xxxx xx, 2003}\accepted{xxxx xx, 2003}
\abstract{We present the first measurement of the angular two-point 
correlation function for AKARI 90-$\mu$m point sources, detected 
outside of the Milky Way plane and other regions characterized by 
high Galactic extinction, and categorized as extragalactic sources 
according to our far-infrared-color based criterion (Pollo et al. 2010). 
This is the first measurement of the large-scale angular clustering of galaxies 
selected in the far-infrared after IRAS measurements. Although a full 
description of clustering properties of these galaxies will be obtained 
by more detailed studies, using either spatial correlation function, 
or better information about properties and at least photometric redshifts 
of these galaxies, the angular correlation function remains the first 
diagnostics to establish the clustering properties of the catalog and
observed galaxy population. We find a non-zero clustering signal 
in both hemispheres extending up to $\sim 40$ degrees, without any
significant fluctuations at larger scales. 
The observed correlation function is well fitted by a power law function. 
The notable differences
between a northern and southern hemisphere are found, which can be probably
attributed to the photometry problems and point out to a necessity
of performing a better calibration in the data from southern
hemisphere.
}

\begin{document}
\label{firstpage}
\maketitle
\copyrighttext{}

\section{Introduction}

According to the now widely accepted paradigm of the gravitational instability theory, 
galaxies formed and evolved inside dark matter halos. 
These haloes grew and merged under the effect of gravity, starting from 
the primordial almost homogeneous distribution, which is imprinted in the 
cosmic microwave background see, e.g., White and Rees (1978). 
Analysis of the galaxy clustering is then believed to be the key to understand 
the evolution of the underlying dark matter field, and hence the Universe itself. 
It is therefore an important issue to understand the bias between the distribution 
of galaxies and the underlying dark matter density field, and how it depends 
on galaxy properties. 

The Infrared Astronomical Satellite (IRAS: Neugebauer et al.\ 1984) has brought 
a great amount of statistics.
Especially in cosmology, the IRAS Point Source Catalog (PSC) provided a great homogeneous dataset 
of galaxies which has driven statistical studies drastically.
A vast number of studies have been done based on IRAS galaxies.
Early studies were based on the angular correlation (Rowan-Robinson and Needham 1986; Lahav et al.\ 1990; Babul and Postman 1990,
Liu et al.\ 1994).
Later, thanks to various IRAS redshift surveys (e.g., Rowan-Robinson et al.\ 1991; Strauss et al.\ 1992; 
Fisher et al.\ 1995; Saunders et al.\ 1990), 
tremendous progress has been brought in the spatial distribution or
correlation function analysis 
(e.g., Efstathiou et al.\ 1990; Saunders et al.\ 1992; Hamilton 1993; Fisher et al.\ 1994; Peacock et al.\ 1997).
In these studies, IR galaxies were used as a tracer of mass distribution in galaxies, explicitly
or implicitly.

However, this might not be regarded as an appropriate assumption anymore, 
since it was found that the amount of dust in galaxies is not strongly correlated to the stellar mass
(e.g., Iyengar et al.\ 1985; Tomita et al.\ 1996).
Indeed, a significant relative bias of IRAS galaxies to optical ones was found 
(e.g., Babul and Postman 1990; Lahav et al.\ 1990; Peacock et al.\ 1994).
Nowadays, the IR emission from galaxies is known to be a good
tracer of star formation activity, especially for actively star-forming 
galaxies, through the heating of dust grains by OB stars 
(e.g., Buat et al.\ 2007; Takeuchi et al.\ 2010; Murphy et al.\ 2011).
Therefore, in a modern context, the large-scale structure of dusty galaxies
is regarded as the star-formation density field in galaxies, which may be
important to connect the dark matter field and star formation activity
(e.g., Malek et al.\ 2010; Amblard et al.\ 2011).

This view has been supported by a vast number of analysis of clustering 
of infrared galaxies at scales up to a few degrees in various surveys.
After IRAS, IR correlation function have been mainly estimated based on 
deep surveys.
Gonzalez-Solares et al.\ (2004) estimated the angular correlation function of
ISO $15\mbox{-}\mu$m galaxies in the European Large-Area Infrared Space
Observatory (ELAIS) S1 survey.
{}From ISO deep surveys, Matsuhara et al.\ (2000) and Lagache et al.\ (2000) performed
a power spectrum analysis of the diffuse FIR background and discovered 
fluctuation due to the large-scale clustering of dusty galaxies.
Subsequently, angular clustering analyses of Spitzer surveys were presented
(e.g., Oliver et al.\ 2004; de la Torre et al.\ 2007; Gilli et al.\ 2007; Magliocchetti et al.\ 2008).
These works were mainly based on MIR data, but thanks to Herschel, 
recently results from longer wavelengths have been gradually reported
(e.g., Maddox et al.\ 2010; Cooray et al.\ 2010; Amblard et al.\ 2011; Magliocchetti et al.\ 2011; 
Planck Collaboration et al. 2011).

After many years since IRAS, the advent of AKARI (ASTRO-F) opened new 
possibilities to explore the whole sky in the far infrared, as a survey-oriented 
space telescope at MIR and FIR (Murakami et al.\ 2007). 
The primary purpose of the AKARI mission is to provide second-generation infrared (IR)
catalogs to obtain a better spatial resolution and a wider spectral coverage than
the IRAS catalog.
All-sky surveys and some pointed deep observations were made by AKARI.
In this work, we present the first measurement of the angular correlation 
function for FIR-selected extragalactic sources from the AKARI All-Sky 
Survey. 
One related work on AKARI Deep Field-South (ADF-S) has been presented 
(Ma{\l}ek et al.\ 2010), but in this work, we made an analysis on much wider area data to
see the large-angle correlation.

This article is organized as follows: in section 2, we present 
the selection of data used for this analysis. In section 3, we present 
and discuss the properties of the angular correlation function of selected
AKARI sources. We conclude in section 4.

\section{Data}

\subsection{AKARI}
AKARI is a Japanese astronomical satellite which was aimed at performing 
various large-area surveys
at the IR wavelengths, from NIR to FIR, with a wavelength coverage of 
$2\mbox{--}160\;\mu$m, as well as
pointed observations.\footnote{Detailed information on the AKARI project, instruments, data and
important results can be found via URL:
http://www.ir.isas.ac.jp/ASTRO-F/index-e.html.}
AKARI is equipped with a cryogenically cooled telescope of 68.5~cm aperture diameter and two
scientific instruments, the Far-Infrared Surveyor (FIS; Kawada et al.\ 2007) and the
Infrared Camera (IRC; Onaka et al.\ 2007).

\subsection{AKARI all-sky surveys}

Among most significant astronomical observations performed by AKARI, 
an all sky survey with FIS and IRC has been carried out; 
it is referred to as the AKARI All-Sky Survey. It is the second
ever performed all sky survey at FIR, after IRAS.
The FIS scanned 96 \% of the entire sky more than twice in the 16 months of
the cryogenic mission phase. 
In March 2010, the AKARI/FIS Bright Source Catalogue v.1.0 has been released to the
scientific community. It contains in total 427 071 point sources
measured at 65, 90, 140, 160 $\mu$m. Hereafter, we use a notation 
$S_{65}$, $S_{90}$, $S_{140}$ and $S_{160}$ for
flux densities in these bands.

The position accuracy of the FIS
sources is $8''$, since the source extraction is made
with grids of this size. Effective size of the point spread function 
of AKARI FIS in FWHM is estimated to be $37 \pm 1''$,
$39 \pm 1$, $58 \pm 3''$, and  $61 \pm 4''$ at 65 $\mu$m, 90 $\mu$m, 
140 $\mu$m, and 160 $\mu$m, respectively \citep{kawada07}.
Errors are not estimated for each individual source, but instead 
they are in total estimated to be 35~\%, 30~\%, 60~\%, and 60~\% at 
65 $\mu$m, 90 $\mu$m, 140 $\mu$m, and 160 $\mu$m, respectively 
\citep{yamamura10}.

Since FIS has sensitivity at longer wavelengths than IRAS, we can expect 
a different composition of sources: we should see objects with cool 
dust which were difficult to detect by IRAS bands. Consequently,
the clustering properties of galaxies selected from the AKARI FIS catalogs
can also be different from the corresponding IRAS galaxies.

\subsection{Selection of Extragalactic Sources}

As mentioned before, the complete FIS All-Sky Survey contains 
427 071 point sources. Since the primary detection 
was performed at 90 $\mu$m, all sources have a flux measurement at least
at this band.

In the further analysis, we restrict ourselves to the area of low
contamination of the Galactic FIR emission ($I_{100} \le 5 \;\mbox{MJy\,sr}^{-1}$)
measured from the Schlegel maps (Schlegel et al.\ 1998), in order to avoid contamination
by sources from Galactic plane and Galactic cirrus emission. 
This procedure excludes also areas of both Magellanic Clouds. 

In order to assure a good quality of AKARI photometric measurements,
we additionally mask the data, restricting ourselves only to the parts
of the sky which were scanned by AKARI at least three times.

This masking procedure leaves us with 13537 sources in the northern hemisphere
and 12096 sources in the southern hemisphere, which gives in total 
25 633 sources.

\begin{figure}[t]
%\centerline{\includegraphics[width=0.45\textwidth]{color_color_ass_selected.ps}} 
\centerline{\includegraphics[width=0.45\textwidth]{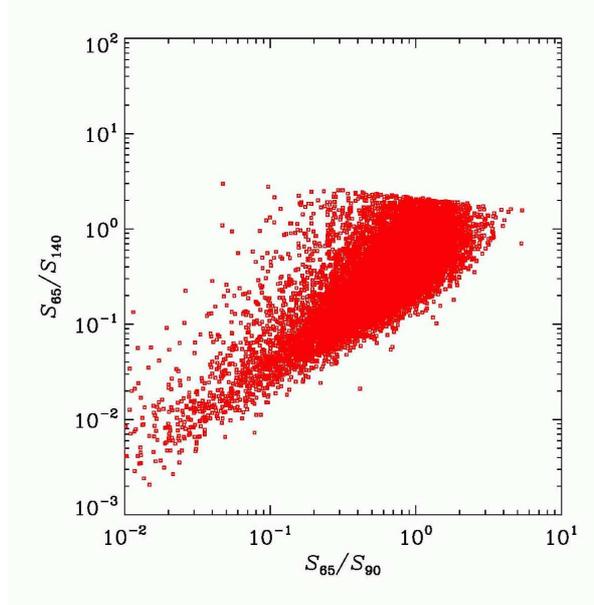}} 
\caption{Color selection of AKARI FIS $90\;\mu$m sources. Galaxy candidates
are presented by dots.}
\label{sel}
\end{figure}

\begin{figure}[t]
\centerline{\includegraphics[angle=90,width=0.45\textwidth]{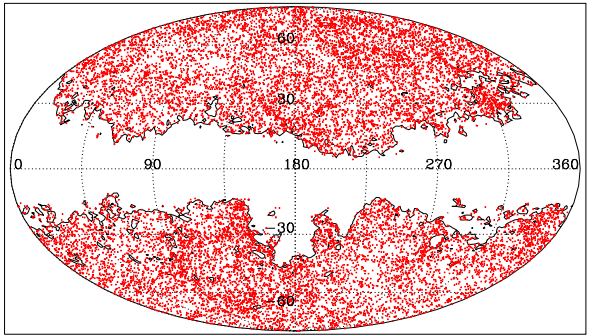}}
\caption{Sky distribution of AKARI FIS galaxies selected by colors.
Only galaxies located on the sky region with $I_{100} < 5\;\mbox{MJy} \, \mbox{sr}^{-1}$
are used in this analysis.}
\label{sky}
\end{figure}

\begin{figure}[t]
\centerline{\includegraphics[width=0.45\textwidth]{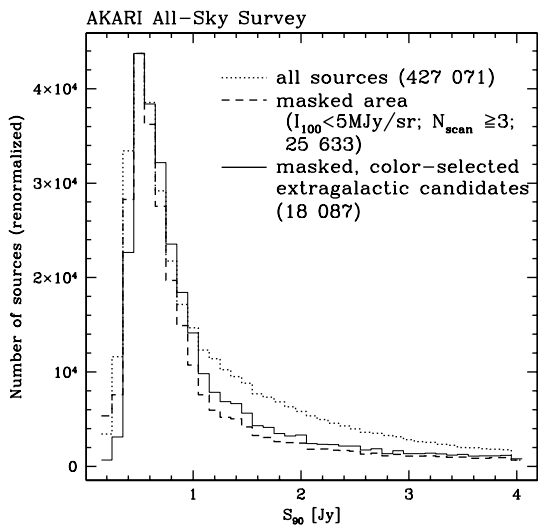}}
\caption{Renormalized histograms of $S_{90}$ fluxes of AKARI FIS sources. 
The dotted line corresponds to the sources from the complete sample. 
Sources from the masked areas, i.e. areas with Galactic cirrus emission 
lower than 5 $\; \mbox{MJy\,sr}^{-1}$ and scanned by AKARI at least three 
times, are denoted by solid line. 
Sources from masked area, selected as galaxy candidates by our far infrared 
color-based criterion, are shown by dashed line.}
\label{hist}
\end{figure}

In Pollo et al.\ (2010), we have presented a method to classify the
AKARI sources in the color-color diagrams only from FIS bands. In order
to be able apply this method to select candidates for extragalactic 
sources in the following analysis, we further restrict ourselves only 
to sources with the full four-band FIS color information, which is 
available for 9700 among already selected sources from the northern 
hemisphere and for 8387 from the southern hemisphere. Then, we applied 
our color-based method to select candidates 
for the extragalactic sources in the low-extinction area. The result
of this selection on the color-color plane
is presented in Figure~\ref{sel}. Sky distribution of all 18 087 sources
left after masking and selection procedures is presented in Figure~\ref{sky}.
This sample is then used for the analysis presented in the next sections.

The results of our selection procedure on $S_{90}$ luminosities of 
the sample can be observed in Figure~\ref{hist} 
which presents the renormalized 
histograms of $S_{90}$ for the complete sample, masked sample and 
masked sample of color-selected galaxies. We can see that the masking 
procedure significantly reduces the bright tail of the distribution
by the removal of the brightest Milky Way sources. 
The color selection
seem to reverse partially this effect, which is a result of the fact
that only sources with full color information, systematically brighter,
were used for this procedure. A few remaining sources with the value 
of $S_{90}$ lower than 0.2, which is below the 3 $\sigma$ detection 
limit of AKARI, were also excluded from the further analysis.

\section{Clustering of AKARI All-Sky Survey Galaxies}

\begin{figure*}[t]
\centerline{
\includegraphics[width=0.45\textwidth,clip]{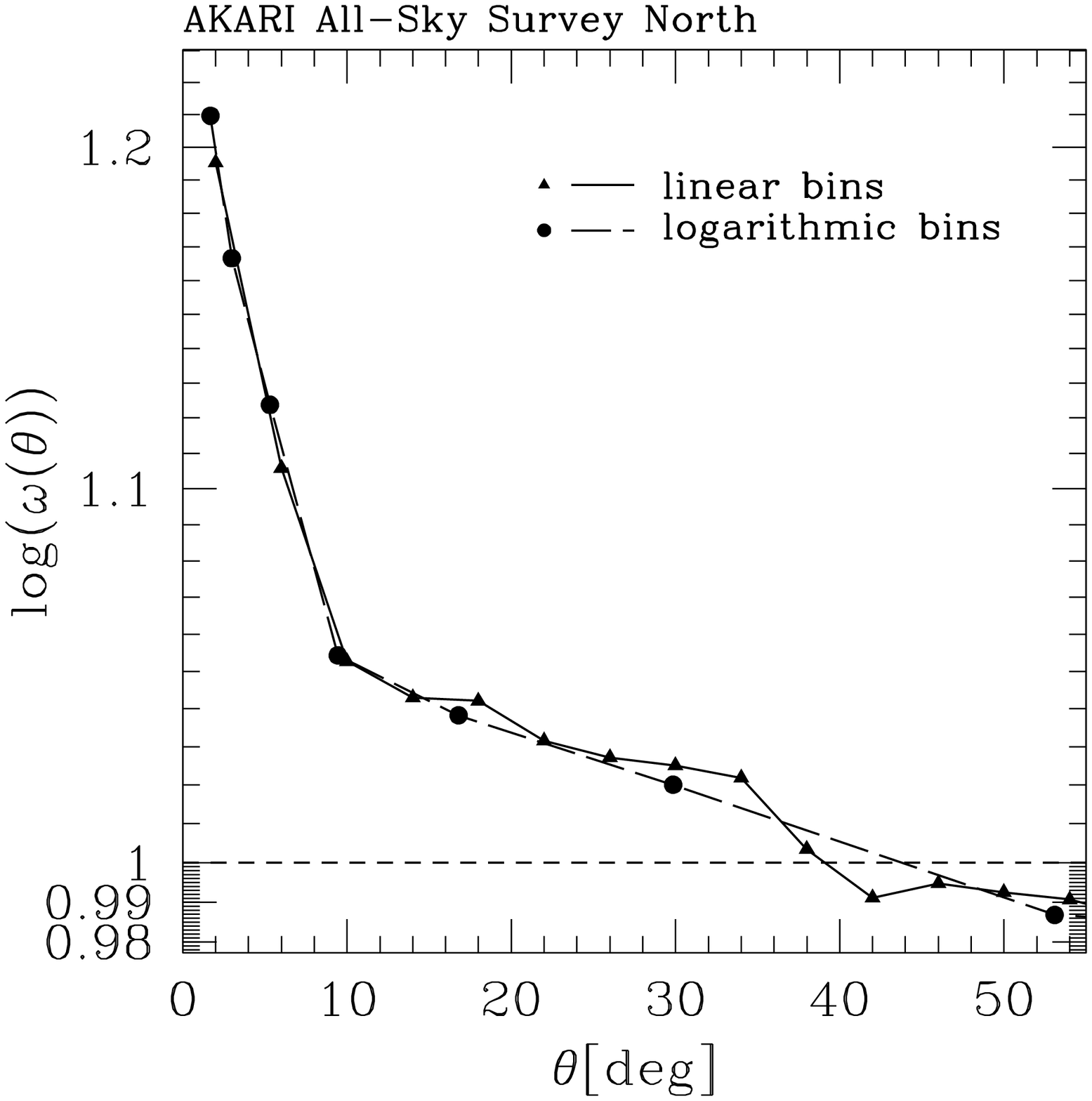}
\includegraphics[width=0.45\textwidth,clip]{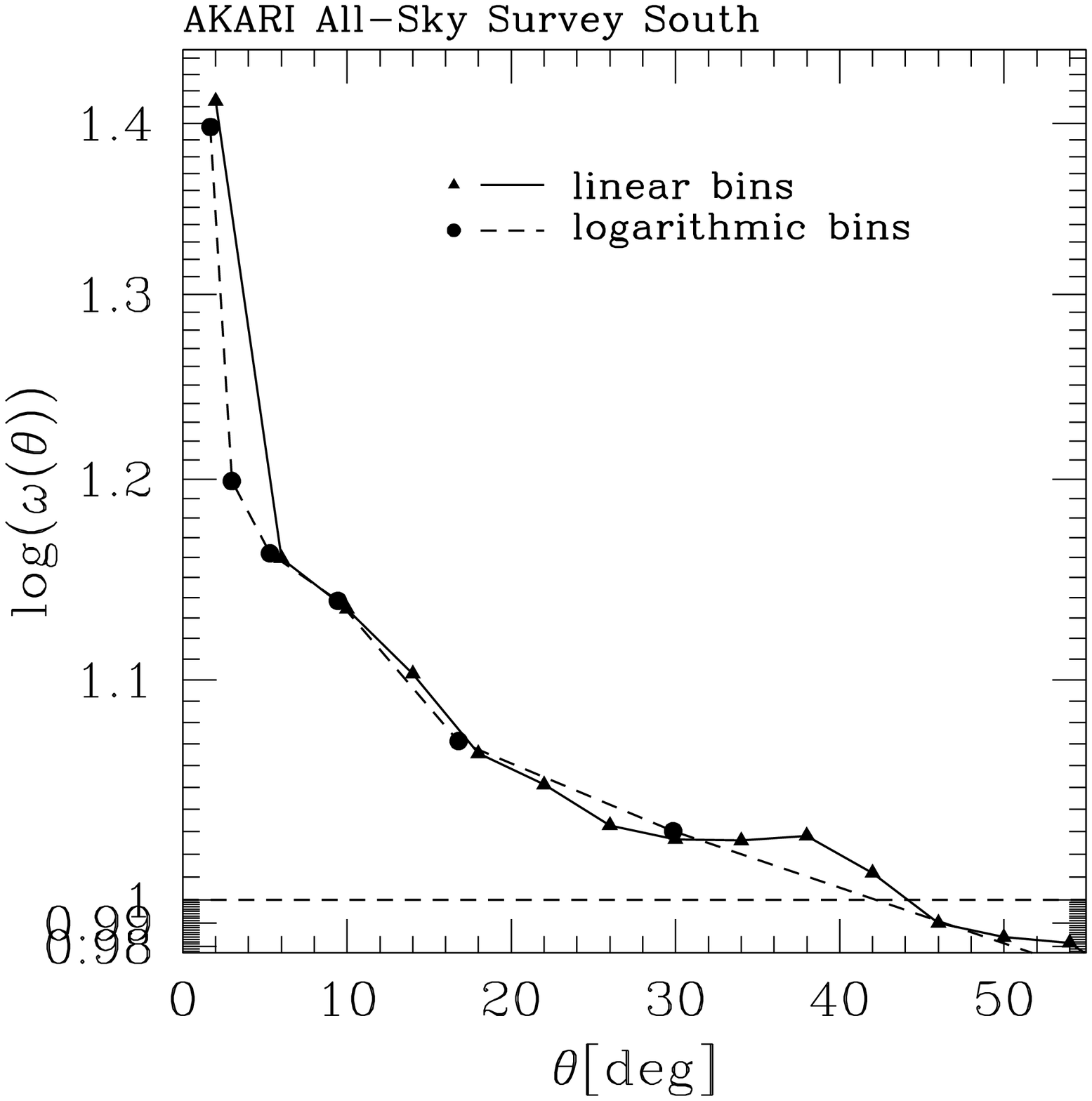}
}
\caption{Angular correlation function, in linear and logarithmic bins,
in the northern (left panel) and southern (right panel) hemisphere
of the AKARI ALL-Sky Survey. In both panels linear bins are marked by
full triangles, connected by solid line, while logarithmic bins are
shown as full circles connected by dashed line. Note a difference
in scale of both panels.}
\label{hs}
\end{figure*}

\subsection{Method}

The two-point angular correlation function, $\omega(\theta)$
is defined as the excess probability above random that a pair 
of galaxies is observed at a given angular separation 
$\theta$ (Peebles 1980).
It is the simplest statistical measurement of clustering, as 
a function of angular scale, and it corresponds to the second 
moment of the distribution. Various recipes, aiming at minimizing
of different sorts of observational biases, have been proposed 
to estimate two-point correlation functions from galaxy surveys. 
In this work, we adopt the angular version of the Landy-Szalay 
estimator (Landy and Szalay 1993), that expresses $\omega(\theta)$ as

\begin{equation}
\omega(\theta) = \frac{N_R(N_R-1)}{N_G(N_G-1)} \frac{GG(\theta)}{RR(\theta)} 
        - \frac{N_R-1}{N_G} \frac{GR(\theta)}{RR(\theta)} + 1 \,\,\,\,\,   .
\label{lseq}
\end{equation}
In this expression, $N_G$ and $N_R$ are the total number (equivalently,
the mean density may be used) of objects respectively in the galaxy sample
and in a catalog of random points distributed within the same survey
volume and with the same photometric mask applied as the one used for
the real data. $GG(\theta)$ is the number of independent galaxy-galaxy 
pairs with separation between $\theta$ and $\theta+d\theta$; 
$RR(\theta)$ is the number of independent random-random pairs within 
the same interval of separations and $GR(\theta)$ represents the number 
of galaxy-random pairs.

%We are aware that the rigorous computation of the whole-sky correlation
%function requires application of a different computational technique,
%based on the decomposition into Legendre polynomials. Commonly used
%estimators, like the one of Landy \& Szalay, are known to be biased
%at larger scales (see e.g. Bernardeau et al. 2002). Even though, they
%are commonly used also for the all-sky correlation measurements ({\bf refs!}).
%The usage of this type of an estimator is then useful to be able to
%compare our results to earlier works, and it makes a reasonably good
%approximation of the real correlation function. It is important, however, 
%that we do rely on the detailed shape of so-obtained correlation function
%on the large scales. 

Different ways of estimating errors on two-point correlation functions 
have been used in the literature (Hamilton 1993; Fisher et al.\ 1994).
Since our aim in this case is the first diagnosis 
of the galaxy clustering in the AKARI data, we do not
apply any refined error estimation method, and we show only the
Poissonian errors. Hence, it should be remembered that they would
indicate the lower limit on the actual errors, since they reflect 
only the information related to the statistical properties of the sample
(see, e.g.,  Fisher et al.\ 1994).

In practice, both spatial and angular correlation function are usually
well fitted by a power-law model:

\begin{equation}
\omega(\theta) = A_w\theta^{1-\gamma},
\label{powlaw}
\end{equation}
with $1-\gamma$ being the slope of the correlation function ($\gamma$ 
itself is then to the slope of a corresponding spatial correlation
function) and $A_w$ is the normalization of the correlation function. 

\subsection{Clustering of Sources in Southern and Northern Hemispheres}

The angular correlation function in the linear scale in shown,
separately for the northern and southern Galactic hemispheres, in the
left and right panels, correspondingly, of Figure~\ref{hs}. In both 
hemispheres we measure a positive signal up to $\theta \sim 40$ degrees. 
For separations larger than $\sim 40$ degrees, the signal remains negative 
without any significant fluctuations. This roughly agrees with the
first clustering measurement for the IRAS sources 
(Rowan-Robinson et al.\ 1986).
In contrast to what was seen in the first IRAS data,
we do not observe any strong difference 
in the shape of the correlation function between northern and southern sky,
in particular between 10 and 40 degrees. However, there are notable 
differences: the most important is that sources in the southern sky
seem to be, at all scales, more strongly clustered than those observed
in the northern sky. 

Since this feature does not correspond to any feature ever measured
in wavelengths, the most probable explanation of this fact would be
an imperfect calibration of photometry of the data from the southern hemisphere, 
possibly being remnants of South Atlantic Anomaly. 
This was also realized in case of the first IRAS data (e.g., Rowan-Robinson and Needham 1986).
Our results indicate, then, that the
All Sky Survey data should be still approached with some caution.

\begin{figure}[t]
\centerline{
\includegraphics[width=0.45\textwidth,clip]{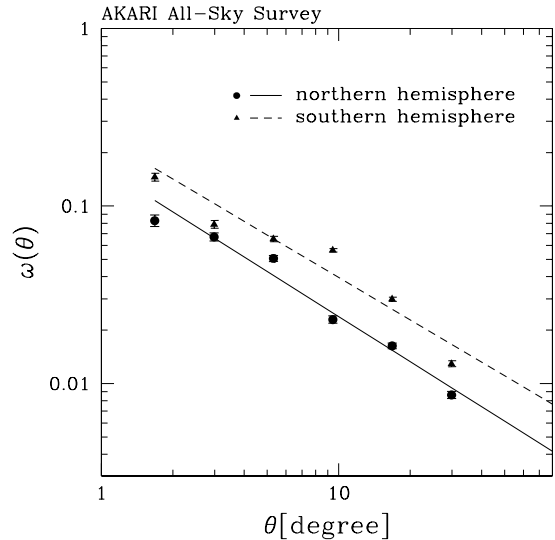}
}
\caption{Power-law fit to the angular correlation function of extragalactic
sources, measured in the northern (full circles, solid line) and southern 
(full triangles, dashed line) hemisphere of the AKARI All-Sky Survey. Points
correspond to the measurements in the logarithmic bins, while lines show
the best power-law fit.}
\label{ns}
\end{figure}

The difference between both hemispheres becomes even clearer
when we a power-law fit to the angular correlation function is made,
as shown in Figure~\ref{ns}. 
Both correlation functions can be fitted
by the power-law function reasonably well on the scale 1-40 degrees, 
but some scale-dependant 
deviations are clearly visible in the function measured in the
southern hemisphere. Both functions have very similar slope 
$\gamma = 1.8 \pm 0.1$, higher than previously measured for these
scales for FIR galaxies.
From this plot it is also well visible that southern galaxies 
seem to be much more strongly clustered than northern ones, with 
the clustering length $A_w = 0.24 \pm 0.01$ degrees, while in the northern
hemisphere we measure $A_w = 0.16^{+0.02}_{-0.01}$.

\begin{figure}[t]
\centerline{
\includegraphics[width=0.45\textwidth,clip]{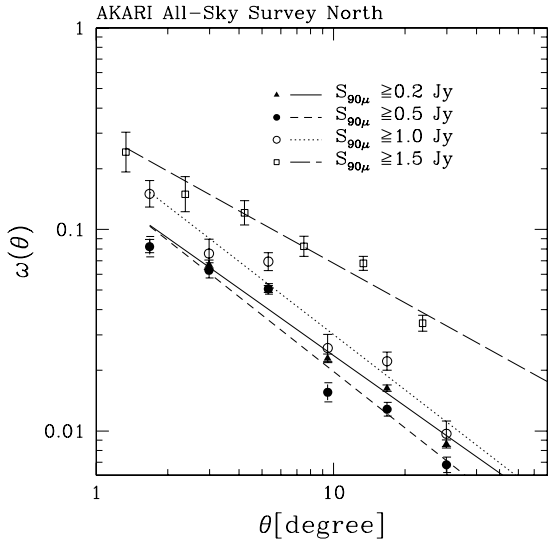}
}
\caption{Angular correlation function in the AKARI All-Sky Survey 
North - comparison of subsamples with different limits of flux density $S_{90}$.  Points
correspond to the measurements in the logarithmic bins and lines show
the best power-law fit. Full triangles and solid line correspond to 
the whole sample, i.e. $S_{90} > 0.2$ Jy. Full circles and short-dashed 
line correspond to the sample with a limit $S_{90} > 0.5$ Jy. Open circles 
and dotted line correspond to the sample limited by $S_{90} > 1$ Jy. 
The brightest sample $S_{90} > 1.5$ Jy is shown by open squares and 
long-dashed line. }
\label{cor_lum}
\end{figure}

\subsection{Flux density dependence of clustering in the AKARI All-Sky Survey North}

Dependence of clustering properties of AKARI FIS galaxies on their 
flux density in 90 $\mu$m is presented in Figure~\ref{cor_lum} and
Table 1. A general
trend is in the agreement with the behavior expected from the hierarchical
model of structure formation, and with other similar measurements: 
brighter galaxies are clustered stronger than fainter ones, and the 
clustering length rises with the limiting flux density. The reversal
of this trend can be observed in case of the two faintest samples:
galaxies with $S_{90} > 0.5$ Jy seem to be less clustered than the
complete sample limited by $S_{90} > 0.2$ Jy. This latter result
might indicate that the photometric measurements are systematically
biased for some of the faintest sources. The clustering of the 
brightest subsample is the strongest, and the best-fitted slope
is less steep than in case of fainter sources, which is more similar
to other far-infrared surveys.

\begin{table*}[t] % two-column -> {table*}
\renewcommand{\arraystretch}{1.2}
\vspace{-.3cm}
\caption{Clustering properties of four subsamples with different flux density 
$S_{90}$ in the AKARI All-Sky Survey North.}
\vspace{-.1cm}
\begin{center}
\begin{tabular}{cccc} \hline
limiting $S_{90}$ [Jy]& Number of galaxies & $A_w$ [deg] & $\gamma$ \\ \hline
0.2 & 8472 & $0.16^{+0.02}_{-0.01}$ & $1.8\pm0.1$ \\ \hline
0.5 & 5493 & $0.16\pm0.02$          & $1.9\pm0.1$ \\ \hline
1.0 & 2233 & $0.24^{+0.05}_{-0.04}$ & $1.9\pm0.2$ \\ \hline
1.5 & 1282 & $0.30^{+0.07}_{-0.05}$ & $1.7^{+0.1}_{-0.1}$ \\ \hline
\end{tabular}
\end{center}
\end{table*}% two-column -> {table*}

\section{Summary and conclusions}

We present the first measurement large-scale clustering of the 
far-infrared galaxies in the AKARI FIS All-Sky Survey. 
We have measured the angular two-point correlation function for 
the 90~$\mu$m-selected sample of galaxies in the northern 
and southern hemisphere. 
Our conclusions are as follows:
\begin{enumerate}
\item We find a positive signal up to 
$\sim$ 40 degrees, in all scales between 1 and 40 degrees
reasonably well fitted by a single power-law function with 
$\gamma \sim 1.8\pm 0.1$, and the amplitude $A_w = 0.16^{+0.02}_{-0.01}$ for
the northern and $0.24\pm 0.01$ for the southern hemisphere. 
\item We suggest that this north-south difference might be a result of 
calibration problems in the data due to the southern hemisphere, possibly 
related to the South Atlantic Anomaly. 
\item We observe the increase of clustering
length with increasing flux density limit of the sources, in accordance with 
expectations for a sample of relatively nearby galaxies.
\end{enumerate}
This measurement of clustering for dusty
galaxies will make it possible to relate the density field of galaxies with hidden 
strong star-forming activity to the general population
of galaxies, i.e., the relative bias of dusty star-forming galaxies.

\acknowledgments
This work is based on observations with AKARI, a JAXA project with the participation of ESA. 
AP has been supported by the research grant of the Polish Ministry
of Science Nr N N203 51 29 38. 
TTT has been supported by the Grant-in-Aid for the Scientific
Research Fund (20740105, 23340046) commissioned by the MEXT. 
TTT, TLS, and SO have also been partially supported from the Grand-in-Aid for the Global
COE Program ``Quest for Fundamental Principles in the Universe: from
Particles to the Solar System and the Cosmos'' from the Ministry of
 Education, Culture, Sports, Science and Technology (MEXT) of Japan.
%%\lastpagecontrol{20cm}

\email{A. Pollo (e-mail: apollo@fuw.edu.pl)}
\label{finalpage}
\lastpagesettings
\end{document}